\documentclass[aps,prb,showpacs,twocolumn,superscriptaddress]{revtex4}
\usepackage{amsmath,amsfonts,amssymb}
\usepackage[dvips]{graphicx}

\input{epsf}

\renewcommand{\Im}{\mathop{\rm Im}\nolimits}
\renewcommand{\Re}{\mathop{\rm Re}\nolimits}

\newcommand{\ds}{\displaystyle}
\newcommand{\uu}{\uparrow\!\uparrow}
\newcommand{\dd}{\downarrow\!\downarrow}
\newcommand{\ud}{\uparrow\!\downarrow}
\newcommand{\du}{\downarrow\!\uparrow}
\newcommand{\vsigma}{\mbox{\boldmath $\sigma$}}

\newcommand{\vpi}{\mbox{\boldmath $\pi$}}

\newcommand{\Tr}{\mathop{\rm Tr}\nolimits}
\newcommand{\Sp}{\mathop{\rm Sp}\nolimits}

\begin{document}

\title{Diffusive magnetotransport in a $2D$ Rashba system}

\author
{S.G.~Novokshonov\,$_{}^{1}$, and A.G.~Groshev\,$_{}^{2}$\\
$_{}^{1}$\,Institute of Metal Physics, Ural Division of RAS, Ekaterinburg,
Russia\\
$_{}^{2}$\,Physical-Technical Institute, Ural Division of RAS, Izhevsk,
Russia}

\date{\today}


\begin{abstract}
We present calculations of the conductivity tensor $\sigma$ of a $2D$--system
with the Rashba spin--orbit interaction (SOI) in an orthogonal magnetic field, 
with allowance for electron elastic scattering by a Gaussian $\delta$--correlated 
random potential in the self--consistent Born approximation. The calculations are 
performed proceeding from the Kubo formula using a new exact representation of 
the one--particle Green function of the $2D$--system with SOI in an arbitrary 
magnetic field. We have obtained the analytical expressions for the density of 
states and $\sigma$ which have a simple interpretation in terms of the two--subband 
model and hold good in a wide range from the classical magnetic fields 
$(\omega_{c}^{}\tau\ll 1)$ up to the quantizing ones $(\omega_{c}^{}\tau\gtrsim 1)$. 
The numerical analysis of the Shubnikov --- de Haas oscillations of the kinetic 
coeffitients and of their behavior in the classical fields region is performed.

\end{abstract}

\maketitle

\section{Introduction}

The growing interest in studying the spin--orbit interaction (SOI) in 
semiconductor $2D$--structures is mostly due to its potential application
to the spin--based electronic devices \cite{zutic_etal_2004}. There are two
main types of SOI in the quantum well based on zinc--blende--lattice
semiconductors. First, the Dresselhaus interaction \cite{dressel_1955} that
originates from the bulk inversion asymmetry (BIA); second, the Rashba
interaction \cite{rashba} induced by structural inversion asymmetry (SIA)
of the confined field of a quantum well. Both these interactions lead to the
momentum-dependent spin splitting of the electron energy spectrum and to the
formation of quantum states with the hard linked spatial and spin degrees
of freedom of the electrons. They are responsible for many interesting effects
in the transport phenomena like beatings in the Shubnikov --- de Haas (SdH)
oscillations \cite{rashba,luo_etal}; weak antilocalization
\cite{iord_etal_1994,pikus_pikus_1995,knap_etal_1996,golub_2005};
current--induced non--equilibrium spin polarization
\cite{levit_etal_1985,edelst_1990}; spin Hall effect
\cite{dyak_etal_1971,mura_etal_2003}, and so on.

At present there are some sufficiently well developed theories of the kinetic
and spin phenomena in $2D$--systems with SOI in zero or classical weak
$(\omega_{c}^{}\tau\ll 1)$ orthogonal magnetic fields. Here $\omega_{c}^{}=
|e|B/mc$ is the cyclotron frequency, and $\tau$ is the electron scattering time.
As for theoretical studies of the considered systems in strong, and especially
in quantizing $(\omega_{c}^{}\tau\gtrsim 1)$ magnetic fields, there is still no 
satisfactory analytical description of the kinetic phenomena even in the 
usual diffusive regime (without quantum corrections). The complex form of the
eigenspinors and energy spectrum of an electron in the presence of SOI and
a strong magnetic field \cite{rashba} is the main cause of such a situation. 
Direct employment of this basis forces one to proceed almost right 
from the start to the numerical analysis of very cumbersome expressions
\cite{wang_etal_2003,lange_etal_2004,averk_etal_2005,wang_etal_2005}.

The strong magnetic field is however one of the most efficient tools for
investigation of SOI \cite{luo_etal} and manipulation of the spin degrees
of freedom in the semiconductor $2D$--structures. Thus, a rather simple
theoretical description of the kinetic phenomena in the $2D$--systems with
SOI in a strong orthogonal magnetic field becomes a necessity. In the present 
work we consider the problem of calculation of the longitudinal and Hall 
resistances of a disordered Rashba system in the self--consistent Born 
approximation (SCBA).

We have found the {\it exact} relation between the one--particle Green function
(GF) of the Rashba $2D$--electron in an arbitrary orthogonal magnetic field and
the well known GF of an "ideal"\, electron, that is an electon with the ideal
value of the Zeeman coupling $(g_{0}^{}=2)$ and without SOI. This allows one
to obtain the analytical expressions for the density of states (DOS) and the
conductivity tensor $\sigma_{ij}^{}$ in the SCBA. These expressions hold good
in a wide range, from the classically weak magnetic fields
$(\omega_{c}^{}\tau\ll 1)$ up to the quantizing ones $(\omega_{c}^{}\tau
\gtrsim 1)$. They have a simple physical interpretation in the framework
of the two--subband conductor model. On the basis of these results, we perform 
the numerical analysis of the beatings of the SdH oscillations of the considered
kinetic coefficients, as well as of their behavior in the classical magnetic
fields region.

\section{Model}

Let us consider a two--dimensional $(|| OXY)$ degenerate gas of electrons
with effective mass $m$, and effective Zeeman coupling $g$ that moves in a
Gaussian $\delta$--correlated random field $U({\bf r})$ in the presence of
an external orthogonal $({\bf B}|| OZ)$ magnetic field ${\bf B}=\nabla\times
{\bf A}$. We assume the Rashba interaction to be the dominant mechanism of
the energy spin splitting in the absence of a magnetic field. This situation
occurs, for example, in the narrow--gap semiconductor heterostructures, such as
${\rm In\,As}/{\rm Ga\,Sb}$ \cite{luo_etal}, ${\rm In\,Ga\,As}/{\rm In\,Al\,As}$
\cite{nitta_etal}. The one--particle Hamiltonian of the considered system has
the form
\begin{equation}
\label{eq:model_h}
{\cal H}+U=\frac{\vpi_{}^{2}}{2m}+\alpha(\vsigma\times\vpi)\cdot{\bf n}
+\frac{1}{4}g\omega_{c}^{}\sigma_{z}^{}+U({\bf r})
\end{equation}
($\hbar=1$). Here $\vpi={\bf p}-e{\bf A}/c=m{\bf v}$ is the operator of the
kinematic electron momentum; $\vsigma=(\sigma_{x}^{},\,\sigma_{y}^{},\,
\sigma_{z}^{})$ is the vector formed by the Pauli spin matrices; $\alpha$ is 
the Rashba spin--orbit coupling; $g$ is the effective Zeeman coupling
($g$--factor).

In the gauge ${\bf A}=(0,Bx,0)$, the components of the eigenspinors of the
Hamiltonian ${\cal H}$ (\ref{eq:model_h}) of a free ($U({\bf r})=0$) Rashba
electron are expressed through the Landau wave functions $\psi_{n,X}^{}
({\bf r})$ depending on the Landau level number $n=0,1,2,\ldots$ and the 
$X$--coordinate of the cyclotron orbit centre $X=-k_{y}^{}/m\omega_{c}^{}$
\cite{rashba}
\begin{subequations}
\label{eq:r_basis}
\begin{equation}
\label{eq:r_spinor}
\ds\widehat\Psi_{\alpha}^{}({\bf r})=\frac{1}
{\sqrt{1+C_{s,n}^{2}}}\left[\begin{array}{c}
\ds C_{s,n}^{}\psi_{n-1,X}({\bf r})\\[4pt]
\ds\psi_{n,X}^{}({\bf r})\end{array}\right]\,,\quad\alpha=(s,n,X)\,.
\end{equation}
The corresponding energy levels have the following form
\begin{equation}
\label{eq:r_spectrum}
{\cal E}_{s,n}^{}=\left\{\begin{array}{ll}
-\omega_{c}\delta\,,  & n=0,~~s=+1\,,\\
\omega_{c}^{}\big[n+s\sqrt{\delta_{}^{2}+2\gamma_{}^{2}n}\,\big]\,, &
n>0,~~s=\pm 1\,.\end{array}\right.
\end{equation}
\end{subequations}
Here $C_{s,n}^{}=\gamma\sqrt{2n}/\big[s\sqrt{\delta_{}^{2}+2\gamma_{}^{2}n}
-\delta\big]$ is a normalizing coefficient; $\delta=(g-2)/4$ is the relative
deviation of the effective Zeeman coupling from its ideal value $g_{0}^{}=2$
(for definiteness, it is assumed that $\delta<0$ in these equations, but all 
the following results are valid for any sign of $\delta$); and, finally, $\gamma
=\alpha\sqrt{m/\omega_{c}^{}}$ is the dimensionless Rashba spin--orbit coupling.

The quantum number $s=\pm 1$ describes the {\it helicity} of the Rashba
electron eigenstate as in the absence of a magnetic field \cite{edelst_1990}.
Indeed, it can be verified immediately that $s=\pm 1$ is the eigenvalue of the
operator

\begin{equation}
\label{eq:helic_oper}
\nu=\frac{\big[\alpha\vsigma\times\vpi+\omega_{c}^{}\delta\vsigma\big]\cdot
{\bf n}}{\sqrt{2m\alpha_{}^{2}{\cal H}_{0}^{}+
\omega_{c}^{2}\delta_{}^{2}}}\,,
\end{equation}
that is diagonal in the basis (\ref{eq:r_spinor}) and approaches the helicity
operator $(\vsigma\times{\bf p})\cdot{\bf n}/|{\bf p}|$ as $B\to 0$.
Here ${\bf n}$ is the unit normal vector to the considered $2D$--system;
\begin{equation}
\label{eq:ideal_h}
{\cal H}_{0}^{}=\frac{\vpi_{}^{2}}{2m}+\frac{1}{2}\omega_{c}^{}\sigma_{z}^{}
\end{equation}
is the Hamiltonian of the "ideal"\, ($g_{0}^{}=2$) electron in a magnetic field,
which commutes with $\vsigma\cdot{\bf n}$, $(\vsigma\times\vpi)\cdot{\bf n}$,
and with ${\cal H}$ (\ref{eq:model_h}).

In spite of this analogy with the $B=0$ case, we cannot say that the
Rashba electron has in the states (\ref{eq:r_spinor}) the spin projection 
$\pm 1/2$ onto the direction $\alpha\vpi\times{\bf n}+\omega_{c}^{}\delta
{\bf n}$, because the components of the kinematic momentum operator $\vpi$
are not commuting motion integrals. Nevertheless, this interpretation
makes sense in the quasiclassical limit, when one can speak about the electron 
path in a magnetic field. Namely, the quantum number $s=\pm 1$ determines
the value of the spin projection on the instant direction of $\alpha\vpi
\times{\bf n}+\omega_{c}^{}\delta{\bf n}$ that changes along the 
quasiclassical electron path. Thus, the spin configurations of the Rashba
electron states form {\it vortices} in the $XY$--plane with center at the 
origin.

The conductivity tensor $\hat{\sigma}$ of the considered system has just
one independent circularly polarized component $\sigma=\sigma_{xx}^{}+
i\sigma_{yx}^{}$. In the one--electron approximation, it has the form 
\cite{gerhar_1975}
\begin{widetext}
\begin{equation}
\label{eq:kubo}
\sigma=\sigma_{}^{I}+\sigma_{}^{II}=
\frac{e_{}^{2}}{8\pi}\Tr\,V_{+}^{}\!\left[\left.\big[2\Phi_{EE}^{RA}-
\Phi_{EE}^{RR}-\Phi_{EE}^{AA}\big]\right|_{E=E_{F}^{}}^{}
+\!\left.\int_{-\infty}^{E_{F}^{}}\!\big(\partial_{E}-\partial_{E_{}'}\big)
\!\big[\Phi_{EE_{}'}^{AA}-\Phi_{EE_{}'}^{RR}\big]\right|_{E_{}'=E}^{}\!
{\rm d}E\right]\!.
\end{equation}
\end{widetext}
Here, $\Phi_{EE_{}'}^{XY}=\big\langle\hat{G}_{}^{X}(E)V_{-}^{}\hat{G}_{}^{Y}
(E_{}')\big\rangle$ is the current vertex operator; $V_{\pm}^{}=V_{x}^{}\pm
iV_{y}^{}=v_{\pm}^{}\pm 2i\alpha\sigma_{\pm}^{}$ are circularly polarized
components of the full velocity operator [the corresponding components
$\vsigma$ are defined as $\sigma_{\pm}^{}=(\sigma_{x}^{}\pm i\sigma_{y}^{})/2$],
where the last term occurs due to SOI (\ref{eq:model_h}). $\hat{G}_{}^{R(A)}(E)
=1/(E-{\cal H}-U\pm i0)$ is the resolvent (retarded $(R)$ or advanced $(A)$) of
the Hamiltonian (\ref{eq:model_h}), and angular brackets $\langle\ldots\rangle$
denote the averaging over the random field $U$ configurations.

\section{One--electron Green function}

By definition, the one--particle GF is the averaged resolvent of the
Hamiltonian (\ref{eq:model_h}) $\langle\hat{G}_{}^{R(A)}(E)\rangle=\langle
1/(E-{\cal H}-U\pm i0)\rangle$. It is connected with the electron self--energy
operator $\hat{\Sigma}_{}^{R(A)}(E)$ by the relation $(X=R,A)$
\begin{equation}
\label{eq:gf_def}
\langle\hat{G}_{}^{X}(E)\rangle=\left[\begin{array}{cc}
\langle G_{\uu}^{X}(E)\rangle & \langle G_{\ud}^{X}(E)\rangle\\
\langle G_{\du}^{X}(E)\rangle & \langle G_{\dd}^{X}(E)\rangle
\end{array}\right]=\frac{1}{E-{\cal H}-
\hat{\Sigma}_{}^{X}(E)}\,.
\end{equation}

The direct employment of the eigenspinors (\ref{eq:r_basis}) for calculation
of (\ref{eq:gf_def}), or kinetic and thermodynamic properties of the Rashba
system in a strong magnetic field leads to very complicated expressions. One 
is forced almost from the first steps either to turn to numerical calculations
\cite{wang_etal_2003,averk_etal_2005,wang_etal_2005}, or to make simplifying 
approximations like the momentum--independent  spin--splitting energy 
\cite{lange_etal_2004}. This makes more difficult the interpretation of the
results obtained in such a way, as well as the understanding of the whole
physical picture. But it turns out that the GF of the free $(U=0)$ Rashba
system is expressed {\it exactly} through the GF of the "ideal"\, electron in a
magnetic field. This opens up new possibilities for analytical studies of
the considered system. Indeed, it is easy to check that the Hamiltonian of the
free Rashba systems can be presented in the following form
\begin{equation}
\label{eq:h_rd_connect}
{\cal H}={\cal H}_{0}^{}+\nu\sqrt{2m\alpha_{}^{2}{\cal H}_{0}^{}
+\omega_{c}^{2}\delta_{}^{2}}\,.
\end{equation}
Here $\nu$ is the helicity operator defined in Eq.~(\ref{eq:helic_oper}),
${\cal H}_{0}^{}$ is the Hamiltonian of the "ideal"\, electron
(\ref{eq:ideal_h}).

The substitution of the Hamiltonian (\ref{eq:h_rd_connect}) into the resolvent
$\hat{G}(E)=(E-{\cal H})_{}^{-1}$ gives, after some simple algebra, the
following result (here and below, we drop superscripts $R(A)$, if
this does not lead to misunderstundings. Sometimes, for brevity of notations,
we shall not write explicitly the energy arguments of the resolvents or GF's.)
\begin{equation}
\label{eq:step_one}
\hat{G}(E)=\frac{E-{\cal H}_{0}^{}+\big[\alpha(\vpi\times{\bf n})+
\omega_{c}^{}\delta{\bf n}\big]\cdot{\vsigma}}{(E+m\alpha_{}^{2}-
{\cal H}_{0}^{})_{}^{2}-\ds\frac{1}{4}\Omega_{B}^{2}}\,,
\end{equation}
where
\begin{equation}
\label{eq:spin_preces}
\Omega_{B}^{}=2\sqrt{2m\alpha_{}^{2}E+m_{}^{2}\alpha_{}^{4}+\omega_{c}^{2}
\delta_{}^{2}}=\sqrt{\Omega_{}^{2}+4\omega_{c}^{2}\delta_{}^{2}}\,.
\end{equation}
The quantity $\Omega_{B}^{}$ is equal to the magnetic field--dependent frequency 
of the spin precession of the electron with energy $E$ that is responsible for 
the Dyakonov --- Perel spin relaxation mechanism \cite{dyak_etal_1971a}; 
$\Omega$ is the same frequency in the absence of a magnetic field. It should be
noted that the same representation of the one--electron GF can be also obtained 
for a system with the momentum--linear Dresselhaus SOI. For example, in the case 
of a $[001]$--grown quantum well based on the 
${\rm A}_{\rm III}^{}{\rm B}_{\rm V}^{}$ semiconductors, it suffices to replace 
$\vpi\to\tilde{\vpi}=(\pi_{y}^{},\pi_{x}^{})$ in the definition of the helicity 
operator (\ref{eq:helic_oper}), change the sign before the Zeeman term 
($g_{0}^{}=-2$!) in the Hamiltonian of the "ideal"\,electron (\ref{eq:ideal_h}), 
and, finally, to redefine the parameter $\delta\to\delta_{D}^{}=(g+2)/4$. 

The denominator of the right--hand side of Eq.~(\ref{eq:step_one}) depends
on the "ideal"\, electron Hamiltonian alone. Expanding this expression 
into the partial fractions, we obtain the desired representation of the
one--electron GF of the free Rashba system
\begin{eqnarray}
\label{eq:gf_repres}
&\!\!\!\hat{G}&\!\!\!\!(E)=\nonumber\\
&\!\!\!=&\!\!\!\frac{1}{2\Omega_{B}^{}}\sum_{s=\pm 1/2}\frac
{\Omega_{B}^{}+4s\big[m\alpha_{}^{2}-\omega_{c}^{}\delta_{}^{}\sigma_{z}^{}
-\alpha_{}^{}(\vpi\times{\bf n})\cdot\vsigma\big]}{E+m\alpha_{}^{2}
+s\Omega_{B}^{}-{\cal H}_{0}^{}}\nonumber\\
&\!\!\!=&\!\!\!\sum_{s=\pm 1/2}\left[\Phi_{s}^{}-2s\frac{\alpha_{}^{}(\vpi
\times{\bf n})\cdot\vsigma}{\Omega_{B}^{}}\right]\hat{G}(E+m\alpha_{}^{2}
+s\Omega_{B}^{})\,.\nonumber\\
\end{eqnarray}
We use here the same notation ($\hat{G}$) for the GF of the Rashba electron
and for the GF of the "ideal"\, electron. However, this does not lead to
confusion since the latter depends always on the energy arguments like
$E+m\alpha_{}^{2}+s\Omega_{B}^{}$ etc.

It is important that the same representation can be obtained for the averaged
resolvent of the Rashba system in the SCBA. We restrict ourselves here to an
approximation in which the electron self--energy operator is diagonal in the 
spin space. Then, the SCBA equation for $\Sigma_{}^{X}(E)$ has the following
form
\begin{equation}
\label{eq:scba_def}
\hat{\Sigma}(E)=W\langle\Sp\hat{G}(E)\rangle=
\left[\begin{array}{cc}
\Sigma_{\uu}^{}(E) & 0\\
0 & \Sigma_{\dd}^{}(E)\end{array}\right],
\end{equation}
where $\Sp$ denotes the trace only over the spatial degrees of freedom; $W$ is
the mean--square fluctuation of the Gaussian random field ($\langle U({\bf r})
U({\bf r}_{}')\rangle=W\delta({\bf r}-{\bf r}_{}')$). Therefore, it suffices to
make everywhere in Eq.~(\ref{eq:gf_repres}) the following substitutions
\begin{equation}
\label{eq:substitution}
E\,\to\,E-\Sigma_{e}^{}(E)\,\qquad g\omega_{c}^{}\,\to\,g\omega_{c}^{}+
4\Sigma_{o}^{}(E)
\end{equation}
to obtain the desired representations for the averaged GF's in the SCBA. Here
$\Sigma_{e(o)}^{}(E)=\big[\Sigma_{\uu}^{}(E)\pm\Sigma_{\dd}^{}(E)\big]/2$
are the even and odd parts of the electron self--energy. The first
($\Sigma_{e}^{}=\Delta_{e}^{}\pm i/2\tau_{e}^{}$) describes the perturbation
(shift $\Delta_{e}^{}$ and broadening $1/\tau_{e}^{}$) of the one--electron 
energy levels by a random field. The real part of $\Sigma_{o}^{}=
\Delta_{o}^{}\pm i/2\tau_{o}^{}$ defines the renormalization of the Zeeman
coupling (\ref{eq:substitution}), while its imaginary part $\propto 1/
\tau_{o}^{}$ makes a contribution to the overall broadening of the one--electron
energy levels. As a result, we obtain a expression like Eq.~(\ref{eq:gf_repres})
for the averaged GF, where
\begin{equation}
\label{eq:ideal_gf}
\hat{G}_{}^{R(A)}(E+m\alpha_{}^{2}+s\Omega_{B}^{})=\frac{1}{E+m\alpha_{}^{2}
+s\Omega_{B}^{}-{\cal H}_{0}^{}\pm\ds\frac{i}{2\tau_{s}^{}}}
\end{equation} 
is the averaged retarded (advanced) GF of the "ideal"\, electron, and
\begin{eqnarray}
\label{eq:renorm}
\Omega_{B}^{}&\!\!\!=&\!\!\!\frac{1}{2}\big(\Omega_{B}^{R}+\Omega_{B}^{A}\big)
\,,\nonumber\\
\frac{1}{\tau_{s}^{}}&\!\!\!=&\!\!\!\frac{1}{{\tau}_{e}^{}}-is\big(
\Omega_{B}^{R}-\Omega_{B}^{A}\big)=\left(1+s\frac{4m\alpha_{}^{2}}
{\Omega_{B}^{}}\right)\frac{1}{\tau_{e}^{}}+s\frac{4\omega_{c}^{}\delta}
{\Omega_{B}^{}}\frac{1}{\tau_{o}^{}}\nonumber\\
\end{eqnarray}
are the disorder--modified frequency of the spin precession
(\ref{eq:spin_preces}) and the inverse life time of an electron in the $s$--th
spin--split subband. As usual, we do not take explicitly into consideration in
(\ref{eq:ideal_gf}) the one--electron energy levels shift $\Delta_{e}^{}$ that
is absorbed by the normalization condition, but we mean here that the odd shift
$\Delta_{o}^{}$ is included in the definition of the effective $g$--factor in
accordance with (\ref{eq:substitution}). The explicit allowance for the Zeeman
coupling renormalization is particularly important in the SdH oscillations
regime.

\section{Density of states and self--energy}

We first consider the calculation of the DOS $n(E)=\Im\langle\Tr\hat
{G}_{}^{A}(E)\rangle/\pi$ using the above--obtained expression for the 
one--particle GF (\ref{eq:gf_repres}). Here, the symbol $\Tr$ denotes the trace
over the spatial and spin degrees of freedom. For the sake of simplicity, we
shall deal with the case of large filling numbers $(E\gg\omega_{c}^{})$. 
Calculating the trace of resolvent (\ref{eq:gf_repres}) over the spatial and 
spin degrees of freedom, we obtain the following expression for the DOS
\begin{eqnarray}
\label{eq:dos_scba}
n(E)&\!\!\!=&\!\!\!\sum_{s=\pm 1/2}\frac{m_{s}^{}}{m}n_{}^{(0)}\big[E+
m\alpha_{}^{2}+s(\Omega_{B}^{}\pm\omega_{c}^{})\big]\nonumber\\
&\!\!\!=&\!\!\!\sum_{s=\pm 1/2}\frac{m_{s}^{}}{m}n_{s}^{(0)}(E)\,.
\end{eqnarray}
Here, we take into account that the DOS of a spinless electron in an orthogonal
magnetic field $n_{}^{(0)}(E)$ satisfies $n_{}^{(0)}(E)=n_{}^{(0)}
(E\pm\omega_{c}^{})$ at large filling factors $(E\gg\omega_{c}^{})$. The sign
before $\omega_{c}$ is chosen in such a way as to ensure the right--hand limit
$s(\Omega_{B}^{}\pm\omega_{c}^{})\to\pm sg\omega_{c}^{}/2$, as the spin--orbit
coupling approaches zero. The effective mass $m_{s}^{}$ in the $s$--th subband
is defined as
\begin{equation}
\label{eq:eff_mass}
m_{s}^{}=m\left(1+s\frac{4m\alpha_{}^{2}}{\Omega_{B}^{}}\right)=
m\partial_{E}^{}(E+s\Omega_{B}^{})\,.
\end{equation}
In the considered case this expression coincides with the usual definition of
the transport and cyclotron effective masses in the isotropic nonparabolic band
\cite{tsidil_1978}. The symbol $\partial_{E}^{}$ denotes the derivative with 
respect to energy $E$.

In full accordance with the two--subband model, the DOS in
Eq.~(\ref{eq:dos_scba}) is presented as a sum of partial contributions. Using
this expression for the DOS, we can obtain the analytical form of the equation
for the electron concentration $n=\int_{}^{E_{F}^{}}n(E){\rm d}E$ that is the
normalization condition for the Fermi level determination. For example, at $B=0$
we have
\begin{equation}
\label{eq:zero_norm}
n=\frac{m}{\pi}\big(E_{F}^{}+m\alpha_{}^{2}\big)=\frac{m}{\pi}E_{0}^{}\,.
\end{equation}
Thus, the energy $E_{0}^{}=E_{F}^{}+m\alpha_{}^{2}$ corresponds to the Fermi
level in the absence of SOI. Notice that the partial electron concentrations
$n_{s}^{}=m(E_{0}^{}+s\Omega_{B}^{})/2\pi$ depend nonlinearly on the Fermi
energy, in contrast to $n$ (\ref{eq:zero_norm}). Of course, the difference 
between $E_{0}^{}$ and $E_{F}^{}$ is small for weak SOI ($m\alpha_{}^{2}\ll
E_{F}^{}$), but it should be taken into account when analyzing the SdH
oscillations that are very sensitive to the electron spectrum character.

The representation (\ref{eq:dos_scba}) allows one to obtain a simple analytical
experssion for the DOS that holds good up to the quantizing fields region
($\omega_{c}^{}\tau\gtrsim 1$). Indeed, the DOS of a spinless electron in the
large filling factors region ($E\gg\omega_{c}^{}$) has the form
\begin{equation}
\label{eq:dos_spinless}
n_{}^{(0)}(E)=\frac{m}{2\pi}\frac{\sinh\ds\frac{\pi}{\omega_{c}^{}\tau}}
{\cosh\ds\frac{\pi}{\omega_{c}^{}\tau}+\cos 2\pi\ds\frac{E}{\omega_{c}^{}}}
\end{equation}
Inserting Eq.~(\ref{eq:dos_spinless}) into Eq.~(\ref{eq:dos_scba}), we
obtain for the oscillating part of the DOS the following expression
\begin{eqnarray}
\label{eq:dos_sdh}
\Delta n(E_{F}^{})=\frac{2m}{\pi}\exp\left(-\frac{\pi}{\omega_{c}^{}\tau}
\right)\left[\cos 2\pi\frac{E_{0}^{}}{\omega_{c}^{}}\cos\pi\frac{\Omega_{B}^{}}
{\omega_{c}^{}}-\right.\nonumber\\
-\left.\frac{2m\alpha_{}^{2}}{\Omega_{B}^{}}\sin 2\pi\frac{E_{0}^{}}
{\omega_{c}^{}}\sin\pi\frac{\Omega_{B}^{}}{\omega_{c}}\right]
\end{eqnarray}
that is valid in the magnetic fields region under consideration. It follows
from (\ref{eq:dos_sdh}) that energy $E_{0}^{}$ (see Eq.~(\ref{eq:zero_norm}))
defines the main period of the SdH oscillations, and $\Omega_{B}^{}/2$
(see Eq.~(\ref{eq:spin_preces})) defines their beating period that depends
on the magnetic field (see Fig.~1(a)). The second term in Eq.~(\ref{eq:dos_sdh})
appears due to the difference between the effective masses $m_{s}^{}$
(\ref{eq:eff_mass}). In the case of weak SOI ($\Omega\ll E$), the oscillations
of the DOS are determined completely by the first term in Eq.~(\ref{eq:dos_sdh}).
Then, the location of the $k$-th node of beatings is determined by the
condition
\begin{equation}
\label{eq:node}
\omega_{c}^{}=\frac{2\Omega}{\sqrt{(2k+1)_{}^{2}-(g-2)_{}^{2}}}\,.
\end{equation}
This limit was considered in Ref.~\cite{taras_etal_2002}. Unlike the results of
that work, the above--obtained equations still stand in the case of strong
SOI. In addition, we have taken into account the Zeeman splitting of the
electron spectrum that allows to describe more correctly the oscillation
pattern. For example, the Eq.~(\ref{eq:node}) allows to determine both
the spin--orbit $\alpha$ and Zeeman $g$ couplings by measured locations
of two different nodes (see upper curve in  Fig.~1(a)). On the other hand,
the spin precession frequency $\Omega_{B}^{}$ approaches $|\delta|\omega_{c}^{}$
as the magnetic field $B$ increases. Therefore, in this case a gradual
transition from the beatings of the SdH oscillations to the familiar Zeeman
splitting of the oscillating peaks should be observed. The beginning of this
transition can be seen on the lower curve in  Fig.~1(a).

\begin{figure}[t!]
\begin{center}
\includegraphics[scale=0.56]{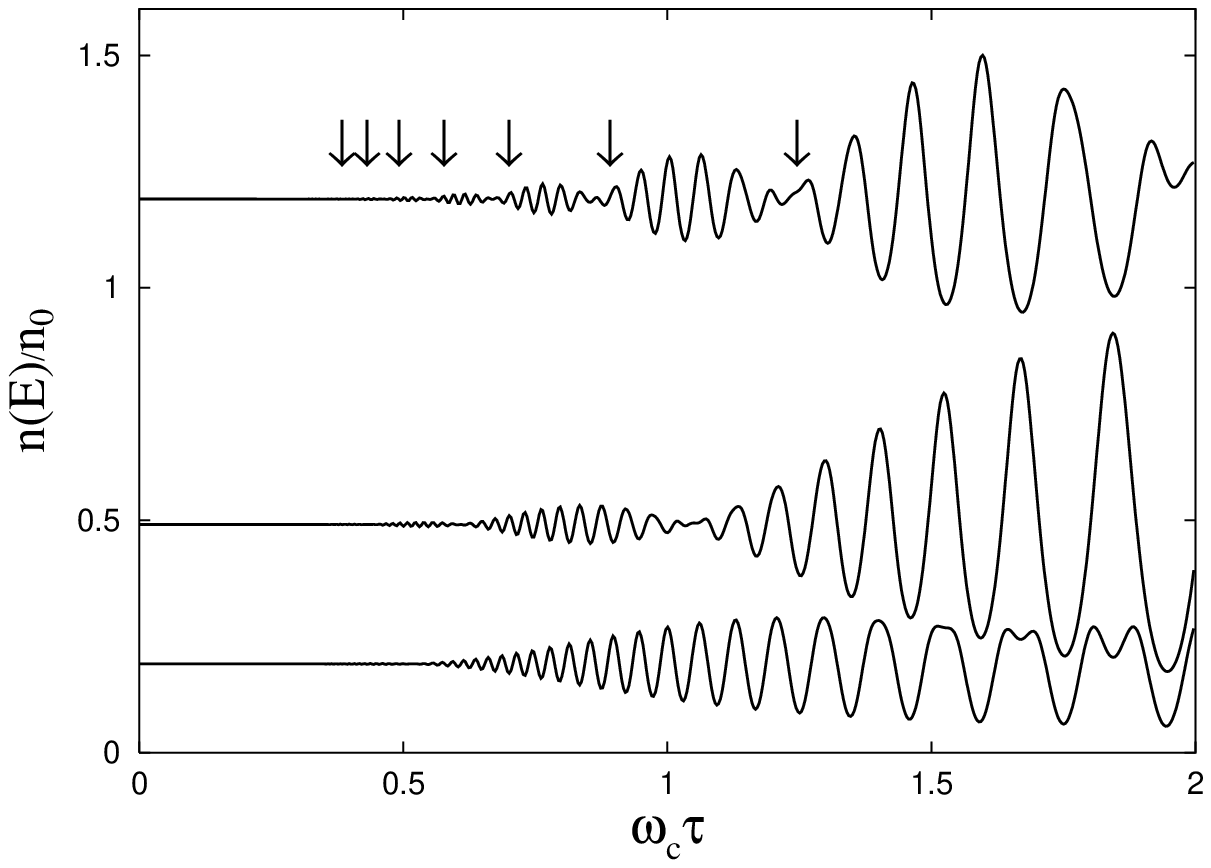}
\includegraphics[scale=0.56]{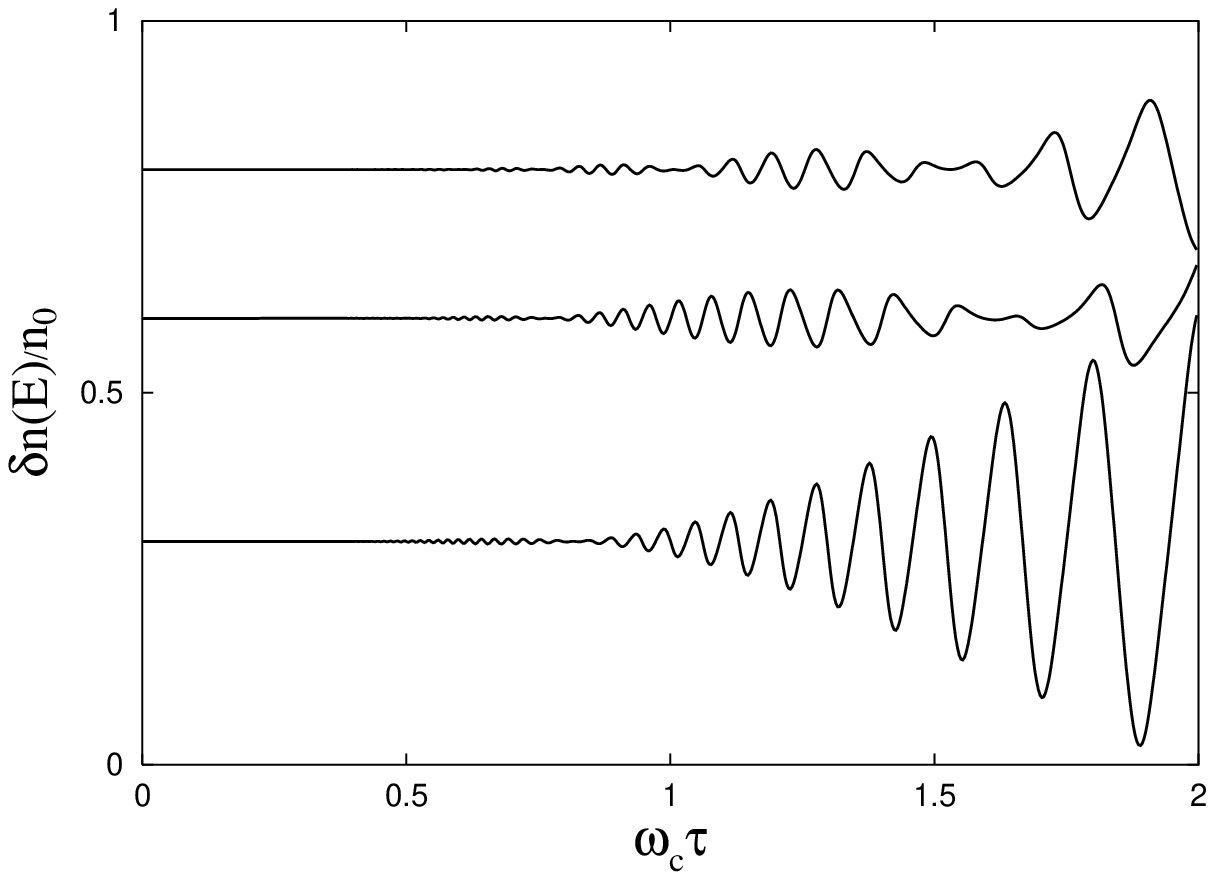}
\caption{Plots of the SdH oscillations of the total DOS (upper panel) and
of the difference of the partial DOS's (bottom panel) of the $2D$ Rashba 
system at fixed $g=2.8$, and $k_{F}^{}l=35.0$, and different 
$\Omega\tau=3.0;\;1.5;\;0.75$ (up to down). The arrows point the nodes 
location that are calculated with Eq.~(\ref{eq:node}).}
\end{center}
\end{figure}

\begin{figure}[t!]
\begin{center}
\includegraphics[scale=0.56]{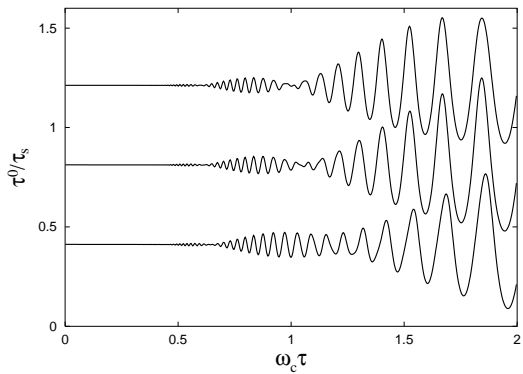}
\caption{Plots of the SdH oscillations of the inverse life time of 
one--electron states in the $s$--th spin--splitted subband at different 
values of Zeeman factor $g=1.8;\;1.0;\;0.2$ (up to down), and fixed values 
of $k_{F}^{}l=35.0$, and $\Omega\tau=1.5$.}
\end{center}
\end{figure}

Another important characteristic of the one--electron states of the $2D$--Rashba
system is the difference of the partial DOS's with opposite spin projections
onto the $OZ$--axis
\begin{equation}
\label{eq:ddos}
\delta n(E)=n_{\uu}^{}(E)-n_{\dd}^{}(E)=\frac{4\omega_{c}^{}\delta}
{\Omega_{B}^{}}\sum_{s=\pm 1/2}sn_{s}^{(0)}(E)
\end{equation}
This quantity is proportional to the derivative of the transverse spin
magnetization with respect to energy $E$ and, therefore, it enters in
the definitition of the effective concentrations of current carriers in
the dissipative part of the $2D$--Rashba system conductivity in an orthogonal
magnetic field (see the next section).

Evidently, the difference of the partial DOS's (\ref{eq:ddos}) vanishes in the
region of classical magnetic fields ($\omega_{c}^{}\tau\ll 1$), but it plays an
important role in the SdH oscillations regime. In the case of large filling
fsctors, the oscillating behavior of this quantity is described by the following
expression
\begin{equation}
\label{eq:ddos_sdh}
\delta n(E_{F}^{})=\frac{2m}{\pi}\frac{2\omega_{c}^{}\delta}{\Omega_{B}^{}}
\exp\left(-\frac{\pi}{\omega_{c}^{}\tau}\right)\sin 2\pi\frac{E_{0}^{}}
{\omega_{c}^{}}\sin\pi\frac{\Omega_{B}^{}}{\omega_{c}^{}}\,.
\end{equation}
Unlike the total DOS (\ref{eq:dos_sdh}), this expression contains just one
oscillating term, because $\delta n(E)$ does not depend on the effective masses
$m_{s}^{}$ (\ref{eq:eff_mass}). Indeed, the difference of the partial
DOS's $\delta n(E)$ is non zero, which is  entirely due to the spin degrees of 
freedom of the electrons. The typical SdH oscillation patterns of $\delta n(E)$
are depicted in Fig.~1(b).

Now, let us turn to the discussion of the electron life time $\tau_{s}^{}$
in the $s$--th spin--split subband which is defined, according to
Eq.~(\ref{eq:renorm}), by the imaginary parts of the even and odd self--energies
$\Sigma_{e(o)}^{}$. In other words, the total life time of the one--electron 
states $\tau_{s}^{}$ is determined by the sum of the weighted relaxation
rates of the orbital and spin degrees of freedom. The first term in this
expression is proportional to the above--considered total DOS, hence its 
magnetic--field dependence coincides up to the scale factor with the patterns
shown in Fig.~1(a). Of particular interest is the last term in
Eq.~(\ref{eq:renorm}) stemming from the Zeeman coupling renormalization. It is
proportional to the difference of the partial DOS's (\ref{eq:ddos}) and,
therefore, plays an important role in the SdH oscillation regime, as shown in
Fig.~2. Notice that the beatings of the SdH oscillations get supressed with the
increase of the relative magnitude of the second term in Eq.~(\ref{eq:renorm}).
Indeed, Eq.~(\ref{eq:node}) determines the location of the beating loops of the 
oscillation instead of the nodes. Thus, the broadening of the Zeeman levels
leads to observable supression of the beatings of the SdH oscillations, as in 
the case of the competition between the Rashba and Dresselhaus SOI's 
\cite{averk_etal_2005}.

\section{Conductivity}

The general expression for the conductivity (\ref{eq:kubo}) consists of two 
different terms. The first of them describes the contribution of the electrons
at the Fermi level, the second one contains the contribitions of all filled
states below the Fermi level. We begin the calculation of the conductivity with
the last term of (\ref{eq:kubo}) $\sigma_{}^{II}$. First of all, it is pure
imaginary and, therefore, makes a contribution in the Hall conductivity alone.
St\u{r}eda \cite{streda_etal} was first to show that, in the absence of SOI,
this part of the conductivity is equal to
\begin{equation}
\label{eq:streda}
\sigma_{}^{II}=i|e|c\left(\frac{\partial n}{\partial B}\right)_{E}^{}\,,
\end{equation}
where $n$ is the electron concentration. It should be pointed out that
Eq.~(\ref{eq:streda}) is {\it exact}, and with the thermodynamic Maxwell
relation $\sigma_{}^{II}$ can be expressed through
$(\partial M/\partial E)_{B}^{}$, where $M$ is the orbital magnetization of the
electron gas. Detailed discussion of $\sigma_{}^{II}$ and its physical
interpretation can be found in survey \cite{pruisken}.

This result is extended immediately to the electron systems with SOI.
Following St\u{r}eda's argument, it can be shown that the part
$\sigma_{}^{II}$ of the of $2D$ Rashba system conductivity is expressed as
\begin{eqnarray}
\label{eq:soi_streda}
\sigma_{}^{II}&\!\!\!=&\!\!\!i|e|c\left[\left(\frac{\partial n}{\partial B}
\right)_{E}^{}-\left(\frac{\partial M_{p}^{}}{\partial E}\right)_{B}^{}\right]
\nonumber\\
&\!\!\!=&\!\!\!
i|e|c\left[\left(\frac{\partial n}{\partial B}\right)_{E}^{}-\frac{g|e|}
{4mc}\big[n_{\uu}^{}(E)-n_{\dd}^{}(E)\big]\right]\nonumber\\
&\!\!\!=&\!\!\!i\frac{|e|nc}{B}\big[n-N_{+}^{}-N_{-}^{}\big]\,,
\end{eqnarray}
where $M_{p}^{}$ is the spin magnetization of the electron gas. The quantities
\begin{equation}
\label{eq:n+-}
N_{s}^{}=\left[E_{0}^{}+s\left(\Omega_{B}^{}+\frac{2\omega_{c}^{2}\delta}
{\Omega_{B}^{}}\right)\right]n_{}^{(0)}\big[E_{0}+s(\Omega_{B}^{}-
\omega_{c}^{})\big]
\end{equation}
($s=\pm 1/2$) are direct analogues of the familiar parameter $n_{\perp}^{}
=En_{}^{(0)}(E)$ that stands for the current carrier concentration in the
dissipative part of the conductivity tensor of spinless $2D$--electrons in
a magnetic field in the SCBA \cite{gerhar_1975}. In the classical magnetic
fields region, they are equal to the partial electron concentrations in the
spin--split bands $n_{s}^{}=m(E_{0}^{}+s\Omega_{B}^{})/2\pi$. On the other
hand, $N_{s}^{}$ (\ref{eq:n+-}) approaches $(E-sg\omega_{c}^{}/2) n_{}^{(0)}
(E-sg\omega_{c}^{}/2)$ in the limit $\alpha\to 0$, in which only the Zeeman 
energy splitting remains. The first two equalities in Eq.~(\ref{eq:soi_streda})
are also {\it exact}. The last one is obtained in the SCBA at large filling
factors ($E\gg\omega_{c}^{}$) using the expression (\ref{eq:dos_scba}) for the
DOS of the $2D$ Rashba system in a magnetic field (for details see Appendix B).
It should be emphasized that a similar SCBA expression is valid also for
spinless electrons \cite{pruisken}.

Now, we turn to the first term in the conductivity (\ref{eq:kubo}). It is quite
easy to show, by identical transformations, that
\begin{widetext}
\begin{equation}
\label{eq:rr_aa_part}
-\frac{e_{}^{2}}{4\pi}\Re\Tr V_{+}^{}\Phi_{EE}^{AA}=
\frac{e_{}^{2}}{2\pi m}\Re\Tr\big\langle\hat{G}_{}^{A}\big\rangle
=\frac{e_{}^{2}}{2\pi m}\sum_{s}\Re\Tr\Big[\Phi_{s}^{R}
\Phi_{s}^{A}\hat{G}_{}^{A}(E_{0}^{}+s\Omega_{B}^{})+
\Phi_{s}^{R}\Phi_{-s}^{A}\hat{G}_{}^{A}(E_{0}^{}-s\Omega_{B}^{})\Big]\,,
\end{equation}
\end{widetext}
where the averaged GF's of the "ideal"\, electron are defined in
Eqs.~(\ref{eq:ideal_gf}) and (\ref{eq:renorm}). In obtaining the last equality
in (\ref{eq:rr_aa_part}), we used the immediately verified identities
\begin{equation}
\label{eq:ident_1}
\Phi_{s}^{R}=\Phi_{s}^{R}\Phi_{s}^{A}+\Phi_{s}^{R}\Phi_{-s}^{A}\,,\qquad
\Phi_{s}^{A}=\Phi_{s}^{R}\Phi_{s}^{A}+\Phi_{-s}^{R}\Phi_{s}^{A}\,.
\end{equation}
The main contribution to the dissipative part of the conductivity is
proportional to the current vertex $\Phi_{EE}^{RA}$ in Eq.~(\ref{eq:kubo}).
If we accept the SCBA (\ref{eq:scba_def}) for the electron self--energy
$\hat\Sigma$, we must evaluate the conductivity (\ref{eq:kubo}) in the ladder
approximation in order to satisfy the particle conservation law. But as show
the calculations, the relative magnitude of the ladder correction to the
conductivity is second--order in the small parameter $1/(k_{F}^{}l)$, and it
can be neglected in comparison with the "bare"\, part of the conductivity
($\Delta\sigma_{}^{\rm lad}/\sigma=0.01\div 0.001$ for typical values of
$k_{F}^{}l=10\div 30$).

Thus, it suffices to calculate the "bare"\, part of the conductivity that
is obtained by replacement $\Tr V_{+}^{}\Phi_{EE}^{RA}\to\Tr V_{+}^{}
\big\langle\hat{G}_{}^{R}\big\rangle V_{-}^{}\big\langle\hat{G}_{}^{A}
\big\rangle$ in the first term in the right--hand side of Eq.~(\ref{eq:kubo}).
We drop the details of calculations that can be found in Appendix~B and proceed
to the results. The overall contribution of the three above--considered parts
has the usual Drude --- Boltzmann form
\begin{equation}
\label{eq:bare_sigma}
\sigma=i\frac{|e|}{B}\left[n-\sum_{s}\frac{\tilde{N}_{s}^{}}{1-i\mu_{s}^{}B}
\right]+\Delta\sigma
\end{equation}
(it is meant here and below that $B\to B/c$). Here, the first two terms
represent the sum of the partial conductivities of the electrons of two
subbands with different  mobilities $\mu_{s}^{}=|e|\tau_{s}^{}/m$, and effective
concentrations
\begin{equation}
\label{eq:new_n+-}
\tilde{N}_{s}^{}=N_{s}^{}
+\frac{4m\alpha_{}^{2}}{\tau_{s}^{}\Omega_{B}^{2}}
\frac{m}{(2\pi)_{}^{2}}(E-s\Omega_{B}^{})\mu_{s}^{2}B_{}^{2}\,.
\end{equation}
This expression differs from $N_{s}^{}$ (see Eq.~(\ref{eq:n+-})) by the
second term that originates from the principal values of the one--electron
GF's.

The last term in Eq.~(\ref{eq:bare_sigma}) represents the small correction to
the conductivity due to the electron intersubband transitions. In the leading
approximation in powers of the smallness parameters $\omega_{c}^{}/E$ and
$\Omega_{B}^{}/E$, it has the form
\begin{equation}
\label{eq:delta_sigma}
\Delta\sigma=-i\frac{|e|}{B}\frac{2m\alpha_{}^{2}}{\Omega_{B}^{2}}\omega_{c}^{2}
\delta\frac{\overline{n_{}^{(0)}}}{(1-i\omega_{c}^{}\tau_{e}^{})_{}^{2}+
\Omega_{B}^{2}\tau_{e}^{2}}\,,
\end{equation}
where $\overline{n_{}^{(0)}}=\sum_{s}n_{s}^{(0)}(E)$ is the DOS at the Fermi
level averaged over the electron subbands. The relative contribution of this
correction to the full conductivity (\ref{eq:bare_sigma}) is of the order of
magnitude $(\omega_{c}^{}/E)_{}^{2}$ and can be neglected in the large filling
factors ($E\gg\omega_{c}^{}$) region.

We emphasize that the contribution to the conductivity from intersubband
transitions vanishes as the magnetic field approaches zero. This would be
expected, because the conductivity tensor in the absence of a magnetic field
is diagonal in the original spin space ($\sigma_{\ud}^{}=\sigma_{\du}^{}
\equiv 0$) by virtue of the momentum parity of the GF's, and the full
conductivity is equal to $\sigma=\sigma_{\uu}^{}+\sigma_{\dd}^{}$
\cite{inoue_etal}. In fact, the case in point concerns the time inversion
symmetry. Using a unitary transformation, it can be turned into a matrix
$s$--representation in which the one--electron GF's are diagonal and, therefore,
$\sigma=\sigma_{\uu}^{}+\sigma_{\dd}^{}=\sigma_{+}^{}+\sigma_{-}^{}$ due to the
trace conservation. Applying an external magnetic field breaks the
above--mentioned symmetry that is responsible for the appearance of the 
intersubband transition--induced conductivity $\Delta\sigma$.

\section{Results and discussion}

First of all, let us summarize briefly the main results obtained in this work.
We have shown that the eigenstates of the $2D$ Rashba electron in an orthogonal
magnetic field are characterized by a special motion integral
(\ref{eq:helic_oper}) that generalizes the notion of {\it helicity}
\cite{edelst_1990}. Using this fact, we have found the relation
(\ref{eq:gf_repres}) between the GF's of the $2D$ Rashba electron and the
"ideal"\, one that holds good for arbitrary orthogonal magnetic fields as well
as for the strong spin--orbit coupling. With the help of this relation, we
have obtained, in contrast to
Refs.~\cite{wang_etal_2003,lange_etal_2004,wang_etal_2005}, the analytical SCBA
expressions for the DOS (\ref{eq:dos_scba}) and magnetoconductivity
(\ref{eq:bare_sigma}) of the $2D$ Rashba system that are valid in a wide range
from the classical magnetic fields up to the quantizing ones ($\omega_{c}^{}\tau
\gtrsim 1$). They permit a simple interpretation in the framework of the model
of two types of current carriers with different concentrations and mobilities.
The spin--orbit as well as the Zeeman splitting of the electron energy are
properly allowed for in these expressions, unlike the results of
Ref.~\cite{taras_etal_2002}. We have shown that the competition of the
relaxation rates of the orbital and spin degrees of freedom in the total inverse
life time $1/\tau_{s}^{}$ of the one--electron states in the $s$--th subband
leads to the supression of beatings of the SdH oscillations as does the
competition of the Rashba and Dresselhaus SOI's \cite{averk_etal_2005}. Finally,
we have shown that the breaking of the time inversion symmetry in a magnetic
field leads to the appearance of the intersubband term in the $2D$ Rashba system
conductivity (\ref{eq:bare_sigma}).

We start the discussion of the results with the conductivity in zero magnetic
field. In this case, it follows immediately from (\ref{eq:bare_sigma}) that
\begin{equation}
\label{eq:zero_cond}
\sigma=|e|(n_{+}^{}\mu_{+}^{}+n_{-}^{}\mu_{-}^{})=\sigma_{D}^{}\left[1-
2\left(\frac{m\alpha}{k_{F}^{}}\right)_{}^{2}\right].
\end{equation}
Thus, the Rashba spin--orbit interaction leads to a decrease in conductivity,
and not to its increase, as it was claimed in Ref.~\cite{inoue_etal}. Let us
note that the authors of that work ignored the difference in mobility between
the electrons of different subbands. As a result, thay obtained a correction
to the conductivity of opposite sign compared to (\ref{eq:zero_cond}), which is  
actually unobservable, because it is absorbed by normalization condition 
(\ref{eq:zero_norm}).

\begin{figure}[t!]
\begin{center}
\includegraphics[scale=0.56]{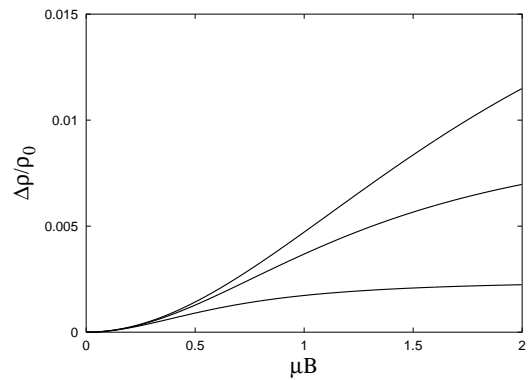}
\caption{The smooth magnetic--field dependence of the resistance calculated 
with Eqs.~\ref{eq:bare_sigma} ---\ref{eq:delta_sigma} at fixed $g=3.5$, and
$k_{F}^{}l=32$, and different $\Omega\tau=1.5;\;1.0;\;0.5$ (up to down).}
\end{center}
\end{figure}

Let us proceed now to the discussion of the magnetotransport in the $2D$ Rashba
system. It is well known that in the two--subband conductors  the classical
positive magnetoresistance and the magnetic--field--dependent Hall coefficient 
are observed (see, for example, \cite{ziman_1972}). The considered system
differs from a classical two--subband conductor in two points. Firstly, the
mobilities and effective concentrations of current carriers (\ref{eq:n+-})
depend on the magnetic field. Secondly, the full conductivity of the $2D$ Rashba
system in an orthogonal magnetic field is not an additive sum of the
intrasubband contributions, but it contains a nonadditive intersubband term
(\ref{eq:bare_sigma}). However, all these factors lead to very slight magnetic
field dependences of kinetic coefficients due to small differences between
concentrations and mobilities of current carriers. For example, the relative
magnitude of the classical magnetoresistive effect comes to only $1\div 2\%$
for typical values of parameters (see Fig.~3).

In discussing the SdH oscillations, we restrict ourselves to the consideration
of the large filling factors $(E\gg\omega_{c}^{})$ region, where the SCBA is
applicable to the description of the one--electron states and kinetic phenomena.
As usual, we extract in the linear approximation the oscillating parts of the 
conductivity that enter through partial DOS's into the effective concentrations
$N_{s}^{}$ (\ref{eq:n+-}) and mobilities $\mu_{s}^{}$. Neglecting the small
differences between concentrations and mobilities of current carriers in the
smooth parts of conductivity, we obtain the expressions for the oscillating
parts of the longitudinal $\rho$ resistance and Hall coefficient $R_{H}^{}$ 
\begin{subequations}
\label{eq:final_sdh}
\begin{eqnarray}
\label{eq:rho_sdh}
\frac{\Delta\rho(B)}{\rho(0)}&\!\!\!=&\!\!\!4\exp\left(-\frac{\pi}
{\omega_{c}^{}\tau}\right)\cos 2\pi\frac{E_{0}^{}}{\omega_{c}^{}}
\cos\pi\frac{\Omega_{B}^{}}{\omega_{c}^{}}\,,\\
\label{eq:rhoh_sdh}
\frac{\Delta R_{H}^{}}{R_{H}^{0}(0)}&\!\!\!=&\!\!\!\frac{2}{\mu_{}^{2}B_{}^{2}}
\exp\left(-\frac{\pi}{\omega_{c}^{}\tau}\right)\cos 2\pi\frac{E_{0}^{}}
{\omega_{c}^{}}\cos\pi\frac{\Omega_{B}^{}}{\omega_{c}^{}}\,.\nonumber\\
\end{eqnarray}
\end{subequations}
Here, $\rho(0)=1/\sigma(0)$ and $R_{H}^{0}\approx-1/|e|nc$ are is the resistance 
(see Eq.~(\ref{eq:zero_cond})) and Hall coefficient in zero magnetic field
respectively. Results of numerical calculations of the SdH oscillations are
performed using total expessions (\ref{eq:bare_sigma})---(\ref{eq:delta_sigma})
are showed in Fig.~4. 

Up to definition of the beatings period, the Eqs.~(\ref{eq:final_sdh}) agree with
expressions for the longitudinal and Hall conductivities obtained in
Ref.~\cite{taras_etal_2002}. As pointed out above, the spin precession
frequency $\Omega_{B}^{}$ (\ref{eq:spin_preces}) and, therefore, beatings period
(see, for example, (\ref{eq:dos_sdh})) depend on the magnetic field due to
allowing for Zeeman coupling. As result, the measurement of two different nodes
location permits to determine both the spin--orbit $\alpha$ and Zeeman $g$
couplings using the Eq.~(\ref{eq:node}).

\begin{figure}[t!]
\begin{center}
\includegraphics[scale=0.56]{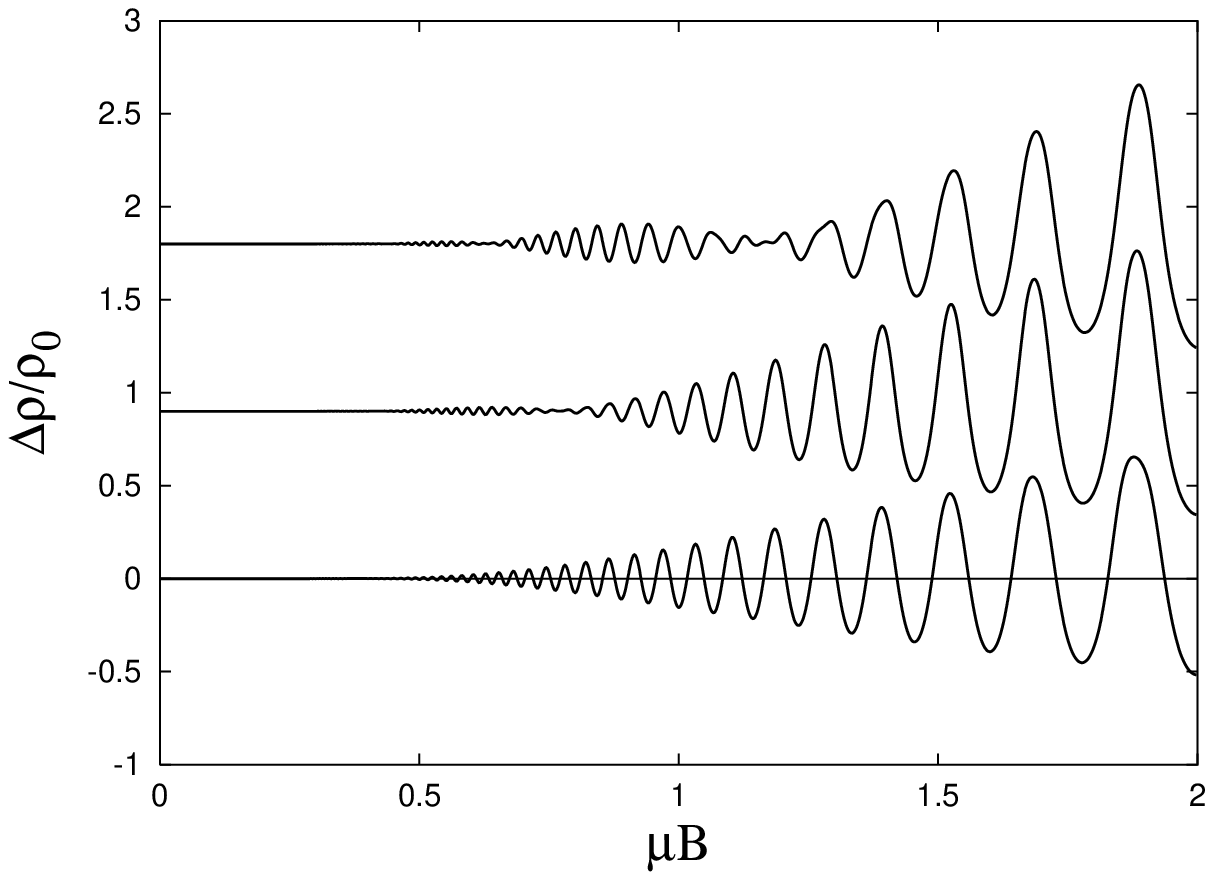}
\includegraphics[scale=0.56]{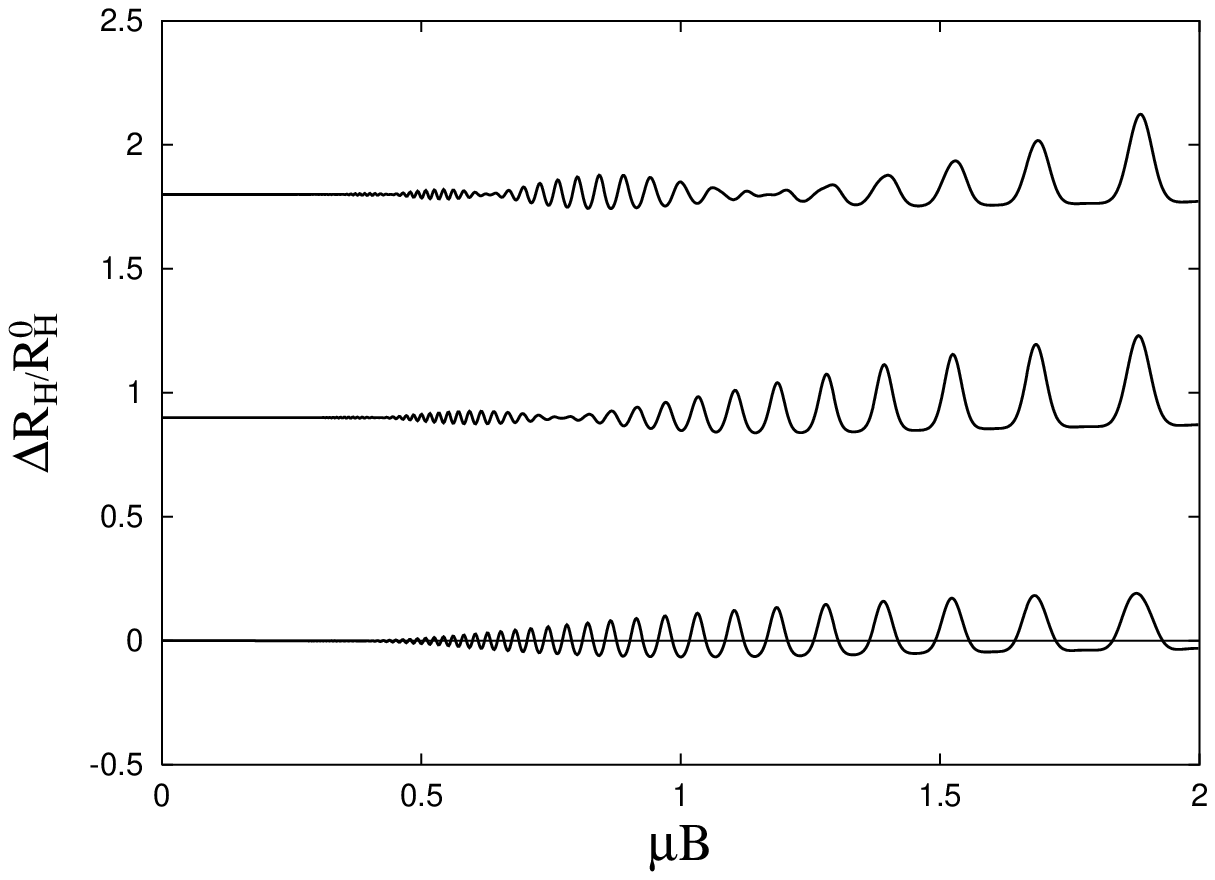}
\caption{Plots of the SdH oscillations of the longitudinal magnetoresistance
(upper panel) and Hall coefficient (bottom panel) of the $2D$ Rashba system
calculated with Eqs.~\ref{eq:bare_sigma} ---\ref{eq:delta_sigma} at
fixed $g=3.5$, and $k_{F}^{}l=32$, and different $\Omega\tau=1.5;\;1.0;\;0.5$
(up to down).}
\end{center}
\end{figure}

\acknowledgments

We thank A.K.~Arzhnikov, A.V.~Germanenko, G.I~Kharus, G.M.~Minkov
and V.I.~Okulov for helpful discussions of results of this work.

This work was supported by the RFBR, grant 04--02--16614

\appendix

\section{Some useful identities}

We obtain here several identities for the off--diagonal matrix elements
$\big\langle G_{\ud(\du)}^{}\big\rangle$ of the one--particle GF
(\ref{eq:gf_def}) that are necessary for calculation of the kinetic coefficients
(see, for example, Eq.~(\ref{eq:bare})).

The matrix of the one--electron GF satisfies the Dyson equation
\begin{equation}
\label{eq:dyson}
\left[E-\Sigma_{e}^{}-\frac{\vpi_{}^{2}}{2m}-\frac{1}{4}g\omega_{c}^{}
\sigma_{z}^{}\right]\langle\hat{G}\rangle-\alpha{\bf n}\cdot
(\vsigma\times\vpi)
\langle\hat{G}\rangle=\hat{I}\,,
\end{equation}
where $\hat{I}$ is the unit $2\times 2$--matrix. It is assumed that the odd
part of the electron self--energy $\Sigma_{o}^{}$ is included into the
effective Zeeman coupling $g$ (see (\ref{eq:substitution})). The $\uu$ matrix
 element of Eq.~(\ref{eq:dyson}) has the form
\begin{eqnarray}
\label{eq:dyson_uu_dd}
\left[E-\Sigma_{e}^{}-\frac{\vpi_{}^{2}}{2m}-\frac{1}{4}g\omega_{c}^{}\right]
\langle G_{\uu}^{}\rangle-i\alpha\pi_{-}^{}\langle G_{\du}^{}\rangle=1\,,
\nonumber\\
\left[E-\Sigma_{e}^{}-\frac{\vpi_{}^{2}}{2m}+\frac{1}{4}g\omega_{c}^{}\right]
\langle G_{\dd}^{}\rangle+i\alpha\pi_{+}^{}\langle G_{\ud}^{}\rangle=1\,.
\end{eqnarray}
Analogous relations can be obtained from the conjugated Dyson equation
(\ref{eq:dyson}). From their comparison with (\ref{eq:dyson_uu_dd}) it follows 
that 
\begin{eqnarray}
\label{eq:ud_du_ident}
\pi_{-}^{}\big\langle G_{\du}^{}\big\rangle=-\big\langle G_{\ud}^{}\big\rangle
\pi_{+}^{}\,,\quad
\pi_{+}^{}\big\langle G_{\ud}^{}\big\rangle=-\big\langle G_{\du}^{}\big\rangle
\pi_{-}^{}\,.
\end{eqnarray}

Now, we replace in (\ref{eq:dyson_uu_dd}) the matrix elements $\langle
G_{\uu(\dd)}^{}\rangle$ with their expressions through the one--particle GF of
a  "ideal"\, electron  (see Eq.~(\ref{eq:gf_repres})). As a result, we have the
following useful relations between $\langle G_{\ud(\du)}^{}\rangle$ and the 
one--particle GF of the "ideal"\, electron
\begin{equation}
\label{eq:r_ud_identity}
\pi_{\mp}^{}\big\langle G_{\du(\ud)}^{}\big\rangle=\pm im\alpha\sum_{s=\pm 1/2}
\frac{E_{0}^{}+s\Omega_{B}^{}}{s\Omega_{B}^{}}G_{\uu(\dd)}^{}(E_{0}^{}+
s\Omega_{B}^{})\,.
\end{equation}
By combining these identities with the corresponding diagonal matrix elements
of Eq.~(\ref{eq:gf_repres}), we obtain
\begin{eqnarray}
\label{eq:r_ud_plus}
\pi_{\mp}^{}\big\langle G_{\du(\ud)}^{}\big\rangle&\!\!\!\mp&\!\!\! 2im\alpha
\big\langle G_{\uu(\dd)}^{}\big\rangle\nonumber\\
=\pm 4im &\!\!\!\alpha &\!\!\!\frac{E\pm\omega_{c}^{}\delta}{\Omega_{B}^{}}\!\!
\sum_{s=\pm 1/2}\!\!sG_{\uu(\dd)}^{}(E_{0}^{}+s\Omega_{B}^{})\,.
\end{eqnarray}

\section{Calculation of conductivity}

We first consider the derivation of the last equality in
Eq.~(\ref{eq:soi_streda}). The immediate differentiation of the electron
concentration with respect to the magnetic field gives the following result
\begin{equation}
\label{eq:streda_1}
\left(\frac{\partial n}{\partial B}\right)_{E}^{}=\frac{n}{B}-\frac{1}{\pi}
\Im\Tr\big\langle\hat{G}_{}^{A}\big\rangle\frac{\partial\cal H}{\partial B}\,.
\end{equation}
Now we should calculate the last term in this expression. Using the
representations (\ref{eq:gf_repres}), (\ref{eq:ideal_gf}) for the averaged GF, 
we write down
\begin{eqnarray}
\label{eq:streda_2}
\Tr\big\langle\hat{G}\big\rangle\frac{\partial\cal H}{\partial B}=\frac{1}{2}
\sum_{s=\pm1/2}\Tr\hat{G}(E_{0}^{}+s\Omega_{B}^{})\left[1+\phantom{\frac{4}{4}}
\right.\nonumber\\
\left.
+4s\frac{m\alpha_{}^{2}-\alpha(\vsigma\times\vpi)\cdot{\bf n}-\omega_{c}^{}
\delta\sigma_{z}^{}}{\Omega_{B}^{}}\right]\frac{\partial\cal H}
{\partial B}\,,
\end{eqnarray}
Let us multiply together the expression in square brackets and the
derivative of Hamiltonian ${\cal H}$ (\ref{eq:helic_oper}),
(\ref{eq:h_rd_connect}), keeping the terms diagonal in the spin space  and 
neglecting the terms linear in $\sigma_{z}^{}$. These latter make contributions  
proportional to $n_{}^{(0)}(E)-n_{}^{(0)}(E\pm\omega_{c}^{})$ and vanish in the 
magnetic field region of interest to us. As a result,  Eq.~(\ref{eq:streda_2})
takes the form
\begin{eqnarray}
\label{eq:streda_3}
\Tr\big\langle\hat{G}\big\rangle\frac{\partial\cal H}{\partial B}=
&\!\!\!\ds\frac{1}{2}&\!\!\!\sum_{s=\pm1/2}\times\nonumber\\
\times\Tr\hat{G}(E_{0}^{} &\!\!\!+&\!\!\! s\Omega_{B}^{})\left[\frac
{\partial{\cal H}_{0}^{}}{\partial B}-\frac{1}{B}\frac{4s}{\Omega_{B}^{}}
\omega_{c}^{2}\delta_{}^{2}\right]\,,
\end{eqnarray}
where ${\cal H}_{0}^{}$ is the Hamiltonian of the  "ideal"\, electron
(\ref{eq:ideal_h}). According to the well known theorem of quantum mechanics,
there is the identity $(\partial H/\partial\lambda)_{nn}^{}=\partial E_{n}^{}/
\partial\lambda$, where $E_{n}^{}$ is the $n$--th eigenvalue of the Hermitian
operator $H$. This make it possible to perform the following substitution
$\partial{\cal H}_{0}^{}/\partial B\to{\cal H}_{0}^{}/ B$ in
Eq.~(\ref{eq:streda_3}). Then, on simple rearrangements, Eq.~(\ref{eq:streda_1})
takes the form 
\begin{eqnarray}
\label{eq:streda_4}
\left(\frac{\partial n}{\partial B}\right)_{E}^{}&\!\!\!=&\!\!\!\frac{n}{B}-
\frac{1}{2\pi B}\sum_{s=\pm 1/2}\times\nonumber\\
&\!\!\!\times &\!\!\!\Im\Tr\left[E_{0}^{}+s\Omega_{B}^{}-\frac{4s}
{\Omega_{B}^{}}\omega_{c}^{2}\delta_{}^{2}\right]\hat{G}_{}^{A}(E_{0}^{}+
s\Omega_{B}^{})\,.\nonumber\\
\end{eqnarray}
Eqs.~(\ref{eq:soi_streda}), (\ref{eq:n+-}) are derived immediately from
this equation. 

Now, let us proceed to the calculation of the dissipative part of the
conductivity that is proportional to $\Tr V_{+}^{}\big\langle\hat{G}_{}^{R}
\big\rangle V_{-}^{}\big\langle\hat{G}_{}^{A}\big\rangle$ in the SCBA.
Performing the trace over the spin degrees of freedom and taking into account
the relations (\ref{eq:ud_du_ident}), we write it down in the form
\begin{widetext}
\begin{eqnarray}
\label{eq:bare}
\Tr V_{+}^{}\big\langle\hat{G}_{}^{R}\big\rangle V_{-}^{}\big\langle
\hat{G}_{}^{A}\big\rangle =\frac{1}{m_{}^{2}}\Sp\Big[\pi_{+}^{}
\big\langle G_{\uu}^{R}\big\rangle\pi_{-}^{}\big\langle G_{\uu}^{A}\big\rangle
+\pi_{+}^{}\big\langle G_{\dd}^{R}\big\rangle\pi_{-}^{}\big\langle
G_{\dd}^{A}\big\rangle-4m_{}^{2}\alpha_{}^{2}
\langle G_{\dd}^{R}\big\rangle\big\langle G_{\uu}^{A}\big\rangle+\nonumber\\
+2\big(\pi_{+}^{}\big\langle G_{\ud}^{R}\big\rangle+2im\alpha\big\langle
G_{\dd}^{R}\big\rangle\big)\big(\pi_{-}^{}\big\langle G_{\du}^{A}\big\rangle
-2im\alpha\big\langle G_{\uu}^{A}\big\rangle\big)\Big]\,.
\end{eqnarray}
Two last terms in the right--hand side of this equation are calculated using 
identities (\ref{eq:gf_repres}), (\ref{eq:r_ud_plus}). We present a more detaled
calculation of one of the two first terms in (\ref{eq:bare}).

For example, let us substitute, in the first term in Eq.~(\ref{eq:bare}), 
the diagonal matrix elements $(\uu)$ of (\ref{eq:gf_repres}) for 
$\big\langle\hat{G}\big\rangle$, and use the identity
\begin{equation}
\label{eq:bare_1}
G_{\uu}^{R}(E_{0}^{}+s\Omega_{B}^{})\pi_{-}^{}G_{\uu}^{A}(E_{0}^{}
+s_{}'\Omega_{B}^{})=\frac{\pi_{-}^{}G_{\uu}^{A}(E_{0}^{}+s_{}'\Omega_{B}^{})-
G_{\uu}^{R}(E_{0}^{}+s\Omega_{B}^{})\pi_{-}^{}}{\ds\frac{i}{\tau_{e}^{}}+
\omega_{c}^{}+s\Omega_{B}^{R}-s_{}'\Omega_{B}^{A}}\,.
\end{equation}

Then, after some simple but cumbersome algebra, we can rewrite the contribution
of this term to the conductivity in the following form
\begin{eqnarray}
\label{eq:bare_2}
\sigma\,\to\,\frac{e_{}^{2}}{\pi m}\sum_{s}\Phi_{s,\uu}^{R}\Phi_{s,\uu}^{A}
\left\{\frac{\left(E_{0}^{}+s\Omega_{B}^{}-\ds\frac{\omega_{c}^{}}{2}\right)
\tau_{s}^{}}{1-i\omega_{c}^{}\tau_{s}^{}}\Im\Sp G_{\uu}^{A}(E_{0}^{}+s
\Omega_{B}^{})-\frac{1}{2}\Re\Sp G_{\uu}^{A}(E_{0}^{}+s\Omega_{B}^{})\big]
\right\}+\nonumber\\
+\frac{e_{}^{2}}{2\pi im}\sum_{s}\Phi_{s,\uu}^{R}\Phi_{-s,\uu}^{A}\left\{
\frac{\left(E_{0}^{}+\ds\frac{s}{2}(\Omega_{B}^{R}-\Omega_{B}^{A})-\ds\frac
{\omega_{c}^{}}{2}\right)\tau_{e}^{}}{1-i(\omega_{c}^{}+2s\Omega_{B}^{})
\tau_{e}^{}}\Sp\big[G_{\uu}^{A}(E_{0}^{}-s\Omega_{B}^{})-G_{\uu}^{R}(E_{0}^{}+
s\Omega_{B}^{})\big]+\right.\nonumber\\
\left.\phantom{\frac{\ds\frac{s}{2}}{C}}+\frac{1}{2i}\Sp\big[G_{\uu}^{A}
(E_{0}^{}-s\Omega_{B}^{})+G_{\uu}^{R}(E_{0}^{}+s\Omega_{B}^{})\big]\right\}\,.
\end{eqnarray}
\end{widetext}
The last terms in curly brackets are cancelled exactly by the corresponding
terms from Eq.~(\ref{eq:rr_aa_part}). The contribution of the second term
from Eq.~(\ref{eq:bare}) can be transformed in a similar way. 

Of course, it is necessary to perform a set of unwieldy some transformations to
obtain  Eqs.~(\ref{eq:bare_sigma}), (\ref{eq:new_n+-}), and
(\ref{eq:delta_sigma}). However, further calculations have a purely
technical character and we omit them.

\end{document}